# New types of nonlinear auto-correlations of bivariate data and their applications

Sanjay Kumar Palit[1*], Sayan Mukherjee[2], D.K.Bhattacharya[3]

[1] Mathematics Department, Calcutta Institute of Engineering and Management,
24/1A Chandi Ghosh Road, Kolkata-700040, INDIA.

[2]Mathematics Department, Shivanath Shastri College, 23/49 Gariahat Road,
Kolkata-700029, INDIA.

[3] Rabindra Bharati University, Kolkata-700050, INDIA.

**Abstract.** The paper introduces new types of nonlinear correlations between bivariate data sets and derives nonlinear auto-correlations on the same data set. These auto-correlations are of different types to match signals with different types of nonlinearities. Examples are cited in all cases to make the definitions meaningful. Next correlogram diagrams are drawn separately in all cases; from these diagrams proper time lags/delays are determined. These give rise to independent coordinates of the attractors. Finally three dimensional attractors are reconstructed in each case separately with the help of these independent coordinates. Moreover for the purpose of making proper distinction between the signals, the attractors so reconstructed are quantified by a new technique called 'ellipsoid fit'.

**Keywords:** chaos, attractor reconstruction, auto-correlation, correlogram.

## 1. Introduction

More than four decades ago, the concept of classical time-continuous chaos was introduced [1]. Since its inception, the interest in this field of research has risen rapidly and several attempts have been made to integrate this fascinating topic [2–15]. However, insights into the experimentally observed irregular behavior of systems are predominantly obtained solely with computer simulations of appropriate Mathematical models. For example we may mention the well known Mathematical model like Lorenz system [16] possessing a unique three dimensional attractor. This approach undermines the trustworthiness of the whole chaos paradigm, since it leads to suspect that chaos is nothing more than a Mathematical artifact, a phenomenon non-existing outside the simulations of the computer. As a remedy to this problem, methods of nonlinear time series analysis were introduced to reconstruct the attractor from the time series itself. Most of these methods were basically developed from the methods of nonlinear dynamics introduced during the late 1970s and 1980s. This reconstruction of the attractor from time series data is useful in order to capture the underlying structure as much as possible. This reconstruction [17-26] of the dynamics of the system on the basis of the given data is of paramount importance, as it ensures that under certain generic conditions such a reconstruction [17-26] is equivalent to the original phase space. This equivalence ensures that differential information is preserved and enables us to proceed hopefully for both qualitative and quantitative analysis. Now we may remember that the fundamental problem in attractor reconstruction [17-26] is to ascertain whether the given time series is chaotic or not. It is known that for a nonlinear time series a positive Lyapunov exponent [27-29] is a strong indicator of chaos. So, first of all we check whether the series is nonlinear or not. In fact, the problem becomes difficult if the series is nonlinear. To test the nonlinearity of a time series, very often the Surrogate data test [30] is used. After checking for nonlinearity and chaotic nature of the signal, we proceed for phase space reconstruction. The most important problem of phase space reconstruction [17-26] is the determination of the time-delay ($\tau$) and the embedding dimension ($m$). As we are interested in geometrical forms of the attractors, so for the sake of visualization we restrict our attractor reconstruction [17-26] to three dimensions only. Hence we do not go into further discussions on how to determine the proper embedding dimension.

[*] Corresponding author. Tel.: +919831004544.
 E-mail address: sanjaypalit@ yahoo.co.in

Now there are several methodologies to determine the proper time-delay. But very often the only measure, which is used in attractor reconstruction [17-26], of real-life problems is either the linear auto-correlation method [31-33] or the Average Mutual Information method [25], the latter being a general measure. However, it should be kept in mind that the proper time-delay for attractor reconstruction must be of moderate magnitude. In fact, it is stated in [34] that if the time-delay is too large, then the chaotic attractors are folded. This may lead to self-intersections of the reconstructed attractors and create a loss of one-to-one property of the reconstruction [17-26]. It is also stated in [34] that if the time-delay is too small, then the reconstructed states do not differ much and the points are scattered around a straight line.

Let us now look back at the actual scenario. Deterministic chaos was observed in many different real-life phenomena that include some Hindustani classical music, some bio-medical signals like ECG [35] and EMG signals, etc. Recently, deterministic chaos was observed in systems as diverse as insects [36-37], reptiles [38] and by laser droplet generation [39]. Most of these real-life signals are non-stationary, and nonlinear with different types of nonlinearities. In case of such time series, we observe that the aforesaid linear measure sometimes gives time-delay, which is too large. In this situation, the notion of Average Mutual Information function [25] is of some help. But we see that same Mutual Information function [25] is used, whatever may be the type of nonlinearity. As a result sometimes we get time-delay that is too small. Hence in both cases, the analysis of the nonlinear time series loses its precision.

Obviously the analysis would be more precise if different types of nonlinear auto-correlations could be chosen to match different types of nonlinearities. These would help in obtaining proper time-delays for the purpose of attractor reconstruction [17-26]. In this connection, it may be mentioned that the new nonlinear auto-correlation is basically taken between a part of the given time series and the corresponding series obtained from curve of best fit of nonlinearity. This is why we prefer to name this process a nonlinear auto-correlation of bivariate data. We are able to show that the value of the time-delay obtained by our newly proposed notion of nonlinear auto-correlation is neither too high nor too small. Consequently our attractor is always found to be better compared to the earlier ones. Naturally the quantification measures on such better forms of three dimensional attractors are expected to distinguish the signals more precisely. As the attractors are now three dimensional, so the earlier method of 'ellipse fit' [40] is no longer applicable. For this purpose, a new type of quantification measure called "ellipsoid fit" is also developed. As an illustration, this newly introduced measure has been applied to obtain proper quantification of the three dimensional attractors.

## 2. Different nonlinear auto-correlations of bivariate data with specific nonlinear trends

In signal analysis, most of the time series encountered in real life have different types of nonlinearities. Since our intention is to find time-delay of moderate magnitude for the purpose of attractor reconstruction [17-26], so without loss of generality we take a small segment (length varies from 150 to 500) of the time signal. This small segment of the signal is then approximated with a nonlinear curve $f$ that gives the best fit. Thus we get two data series, one corresponds to the suitable smaller segment of the time signal and the other one is what is generated through the nonlinear curve $f$.

## 2.1. Some new definitions

**Definition.2.1.1**

Let $X = \{x(k)\}_{k=1}^{N}$ and $Y = \{y(k)\}_{k=1}^{N}$ be two time series and $f(k)$ be the function that approximates $y(k)$ from $x(k)$, $k=1,2,\ldots,N$. Let $Y' = \{f(x(k))\}_{k=1}^{N}$. Then the nonlinear correlation between X and Y, denoted by $r_{X,Y}$ is defined as the linear correlation between $Y = \{y(k)\}_{k=1}^{N}$ and $Y' = \{f(x(k))\}_{k=1}^{N}$, which is given by i.e., 
$$r_{X,Y} = \frac{\sum_{k=1}^{N}\{y(k) - \overline{y(k)}\}\cdot\{f(x(k)) - \overline{f(x(k))}\}}{\sqrt{\sum_{k=1}^{N}\{y(k) - \overline{y(k)}\}^2} \cdot \sqrt{\sum_{k=1}^{N}\{f(x(k)) - \overline{f(x(k))}\}^2}} \quad\ldots\ldots(1)$$

**Definition.2.1.2**

Let $X = \{x(k)\}_{k=1}^{N}$ be a signal and $\{f(k)\}_{k=1}^{N}$ be its best nonlinear curve fit. Then the nonlinear auto-correlation between $X = \{x(k)\}_{k=1}^{N}$ and $Y = \{x(k+m)\}_{k=1}^{N}$ is denoted by $R_X(m)$ and is defined as the linear auto-correlation between $Y = \{x(k+m)\}_{k=1}^{N-m}$ and $Y' = \{f(x(k))\}_{k=1}^{N}$, i.e;

$$R_X(m) = \frac{\sum_{k=1}^{N-m}\{f(x(k)) - \overline{f(x(k))}\}\cdot\{x(k+m) - \overline{x(k+m)}\}}{\sqrt{\sum_{k=1}^{N-m}\{f(x(k)) - \overline{f(x(k))}\}} \cdot \sqrt{\sum_{k=1}^{N-m}\{x(k+m) - \overline{x(k+m)}\}}}, \quad\ldots\ldots(2)$$

where $1 \leq m \leq \dfrac{N}{2}$, if $N$ is even and $1 \leq m \leq \dfrac{N-1}{2}$, if $N$ is odd.

**Remark.2.1.1**
It might appear that the value of auto-correlation would differ significantly if longer part of the signal could be chosen for curve of best fit. But auto-correlation is a ratio. With inclusion of longer part of the signal, both the numerator and the denominator in the expression of auto-correlation change relatively. So the result is self adjusting in nature. Hence it may be stated that there is no loss of generality in taking smaller segment of the signal for suitable curve fit. As a matter of fact, when we consider auto-correlation of a signal, it is basically the correlation between two segments of the signal of finite length differing by one unit only. So there is no need of trying to have nonlinear fit with the whole signal. Again we see that by considering a suitable small segment we ultimately get a time delay, which is not too large or too small and the shape of the attractor also improves. So there is no problem if for accommodating a curve of best fit, we consider that part of the signal which is longest for the purpose. Of course, if no such best nonlinear fit is available for any part of the time signal, then it is not possible to determine the suitable moderate time-delay by using equation (2) for the reconstruction [17-26] purpose. This could happen had the signal been highly non-stationary, because in that case the trend of the signal would vary randomly. But this is a very rare case. So we expect that in most of the cases our new notion of auto-correlation work satisfactorily.

## 3. Determination of nonlinear independent coordinates for attractor reconstruction

We assume that a nonlinear curve of best fit exists for the given time signal. Since the signal is nonlinear, the nonlinear auto-correlation given by (2) is used to determine the auto-correlations between two groups $\{x(k)\}_{k=1}^{l-m}, \{x(k+m)\}_{k=1}^{l-m}$, $150 \leq l \leq 500$, obtained by sub-dividing a smaller segment $\{x_k\}_{k=1}^{l}$ of the time series $\{x_k\}_{k=1}^{N}$. The time-delay $m$ is determined from the two dimensional correlogram diagram [32]. This gives $x(k), x(k+m), x(k+2m)$ $(k=1,2,3,...., N-2m)$ as the independent coordinates for attractor reconstruction [17-26]. Finally the attractor is reconstructed with these independent coordinates.

## 4. Correlogram diagram in two dimensional spaces

For the purpose of reconstruction of the attractor [17-26] in three dimensional spaces, the nonlinear auto-correlation function $R_x(m)$ is calculated for different values of $m$ by using (2). The values of $R_x(m)$ are then plotted against $m$. This is known as Correlogram diagram [32] in two dimensional spaces. Finally, the reconstruction of the attractor [17-26] is carried out by choosing that value of $m$ for which $R_x(m)$ comes nearer to zero for the first time in the two dimensional correlogram diagram [32].

## 5. Attractor reconstruction of Lorenz system by using traditional methods and Newly proposed method of nonlinear auto-correlation

### 5.1. Traditional methods

In order to see that our newly proposed method of nonlinear auto-correlation gives better results compared to the traditional ones, we first consider standard methods for the attractor reconstruction [17-26] of the known Lorenz system [1,16] given by the following differential equations:

$$\frac{dx}{dt} = s(y - x)$$
$$\frac{dy}{dt} = rx - y - xz \quad (3)$$
$$\frac{dz}{dt} = -bz + xy$$

with the initial condition $x(1) = 8$, $y(1) = 9$, $z(1) = 25$ and $s = 10, r = 28, b = \frac{8}{3}$. Its chaotic attractor is given in fig.1.

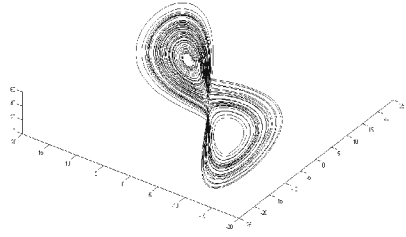

**Fig.1: Attractor of the Lorenz system given by (3).**

Now it is known that the time delay $m$ for its attractor reconstruction [17-26] is found to be $m = 63$ and $m = 16$ measured under standard auto correlation and Average Mutual Information respectively.

The first time delay is not accepted as it is too high. Now for $m = 16$, the attractor is given by fig.2.

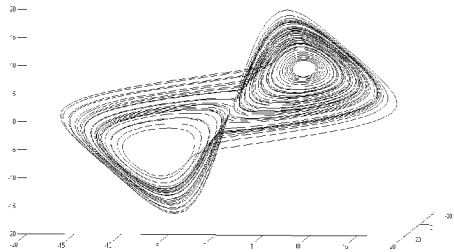

**Fig.2: Reconstructed attractor of the Lorenz system (3) for the time-delay $m = 16$, obtained under the nonlinear measure – Average Mutual Information.**

### 5.2. Newly proposed nonlinear auto correlation methods

We now apply our own methodology [as discussed in section 2] for reconstruction of the attractor [17-26] of the above Lorenz system [16] to a single component $\{x(t)\}_{t=1}^{5000}$ of its solution vector $(x(t), y(t), z(t))$. Obviously, $\{x(t)\}_{t=1}^{5000}$ is nonlinear. For attractor reconstruction [17-26] of Lorenz system, we take a smaller segment of $\{x(t)\}_{t=1}^{5000}$ of length 200 and then find a best possible nonlinear curve fit $f$, which is given by

$$f(t) = a_0 + a_1 \cos(wt) + b_1 \sin(wt) + a_2 \cos(2wt) + b_2 \sin(2wt) + a_3 \cos(3wt) + b_3 \sin(3wt) + a_4 \cos(4wt) + b_4 \sin(4wt) + a_5 \cos(5wt) + b_5 \sin(5wt), \quad \ldots(4)$$

where
$a_0 = 8.44$, $a_1 = 0.05261$, $b_1 = -0.04536$, $a_2 = -0.02356$, $b_2 = -0.09461$, $a_3 = -0.1507$, $b_3 = -0.07338$, $a_4 = -0.4175$, $b_4 = 1.353$, $a_5 = 0.03087$, $b_5 = -0.04881$, $w = 0.02525$.

This is shown in fig.3.

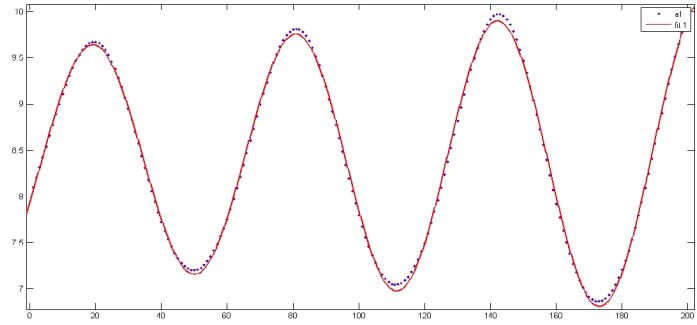

**Fig.3. Best nonlinear curve fit for a smaller segment of the solution component $\{x(t)\}_{t=1}^{5000}$ of Lorenz's system of length 200.**

We then use equation (2) to determine the nonlinear auto-correlation between $X = \{x(t)\}_{t=1}^{200}$ and $Y = \{x(t+m)\}_{t=1}^{200}$, denoted by $R_X(m)$, which is actually the linear auto-correlation [31-33] between $Y = \{x(t+m)\}_{t=1}^{200}$ and $Y' = \{f(x(t))\}_{t=1}^{200}$, where $f$ is the best nonlinear fit available for a smaller part of one of the solution components $\{x(t)\}_{t=1}^{5000}$ of the aforesaid Lorenz system [16]. Thus we obtain $R_X(m)$ for different values of the time-delay $m$. The suitable time-delay $m$ is determined from the two dimensional correlogram diagram [32] given by fig.4.

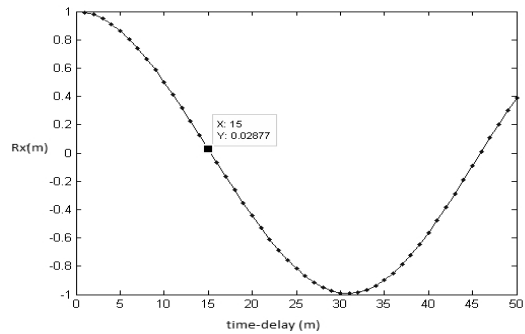

**Fig.4: Two dimensional correlogram diagram (plot of $R_X(m)$ against $m$) for one of the solution component $\{x(t)\}_{t=1}^{5000}$ of the Lorenz system. Here $R_X(m)$ comes nearer to zero for the first time when $m$=15.**

It is evident from the above correlogram diagram that $R_X(m)$ comes nearer to zero for the first time, when $m = 15$. Corresponding to this time-delay the reconstructed attractor is shown in fig.5.

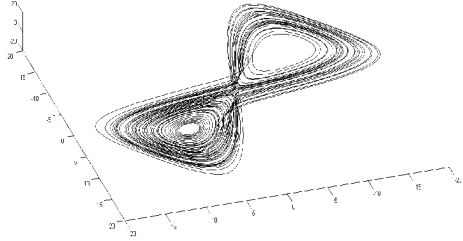

**Fig.5:** Attractor of the Lorenz system (3) reconstructed from the solution component $\{x(t)\}_{t=1}^{5000}$ for the time-delay $m=15$. The independent coordinates for this reconstruction are taken as $x(t), x(t+m), x(t+2m)$.

**Remark 5.1**

If we compare Fig.5 and Fig.2 with Fig.1, we see that Fig.5 is more similar to Fig.1 than Fig.2. Thus it is established that using our new methodology it is possible to reconstruct the attractor of the Lorenz system [1, 16] properly, if not better compared to the traditional ones.

## 6. Attractor reconstruction of some real physical signal with different types of Non-linearity

To carry out the reconstruction of the attractor [17-26], we consider some physical signals like music signals due to three reasons: (1) it is easier for us to collect data of a music signal by available software, (2) music signals are basically nonlinear and are always found to possess different types of nonlinearities, (3) primary detection of differences of the signals can be judged by the musicians. This helps us indirectly to check our scientific predictions about the signals regarding their quantitative differences. Though the signals are basically non-stationary as a whole, yet the smaller segments (length varies between 150 and 500) of those signals seem to have some definite form [41]. Hence there exists best nonlinear curve fits for each of them. But these fitted nonlinear curves are different for different music signals. We consider music signals with their time series plot, their smaller segments along with their best nonlinear curve fits and apply our newly proposed methodology for the reconstruction of the attractors [17-26] for each of them. In each case of such signals we also establish separately that our new methodology is better than the traditional ones.

### 6.1. Music signal of a Hindustani Classical Sarod recital based on raga 'Anandi'

### 6.1.1. Methodology of collecting data

Source- Recorded north Indian Classical Music (Sarod)
Artist- Ustad. Amjad Ali Khan
Rag- Anandi
Software- Adobe Audition 1.5
Samples: 10000
Bits per Samples: 16

Sample rate: 44100
Normalized: False

### 6.1.2. Test of nonlinearity of the signal

The Surrogate data test [30] with $0.01$ significant level and the statistical parameter AMI $(\tau = 1)$ (Average Mutual Information with time-lag 1) of the music signal of raga 'Anandi' is given in Fig.6.

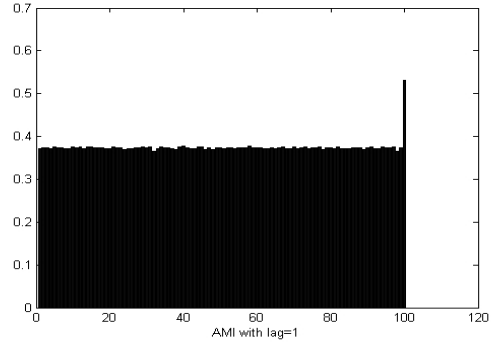

**Fig.6 :** Plot of AMI$_{Anandi}$ and AMI $_{SUR(Anandi)}$ for $\tau = 1$ against signal number, where the first 99 signals are surrogates. Here AMI of the original signal is greater than that of the surrogates. So, the null hypothesis H$_0$ : AMI $_{Anandi}$ ( $\tau = 1$ )=AMI $_{SUR\ (Anandi)}$ ( $\tau = 1$ ) fails in this case.

In this connection, we take the null hypothesis $H_0 : AMI_{Anandi}(\tau = 1) = AMI_{SUR(Anandi)}(\tau = 1)$. If the equality does not hold, we say that null hypothesis fails and alternative $H_A$ holds. From Fig.6, it is seen that in this case the AMI of surrogate data series is not equal to AMI of the given signal. Hence the null hypothesis $H_0 : AMI_{Anandi}(\tau = 1) = AMI_{SUR(Anandi)}(\tau = 1)$ is rejected. Thus, nonlinearity of the present signal of raga 'Anandi' is established through Surrogate data test [30].

### 6.1.3. Test for deterministic chaos

Since the music signal of raga 'Anandi' is found to be nonlinear by Surrogate data test [30], we now test whether the signal possesses chaotic attractor. To do this, we compute Lyapunov exponent $(\lambda)$ [27-29] of the signal. We find that the Lyapunov exponent $(\lambda)$ [27-29] for this music signal is positive, which indicates that the present signal possesses chaotic attractor.

### 6.1.4. Time series plot of the music signal and the best nonlinear curve fit

Let $\{x(k)\}_{k=1}^{10000}$ be 10000 samples of the recorded north Indian Classical Music (Sarod) based on raga 'Anandi'. The time series plot of this signal is given by fig.7.

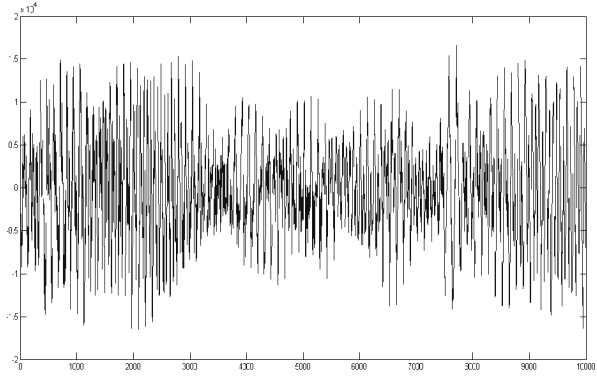

**Fig:7. Time series plot of the music signal based on raga 'Anandi'.**

We next consider a smaller segment of the above music signal of length 200 and find its best non-linear curve fit, which is given by

$$f(t) = a_1 \exp(-((t-b_1)/c_1)^2) + a_2 \exp(-((t-b_2)/c_2)^2) + a_3 \exp(-((t-b_3)/c_3)^2) + a_4 \exp(-((t-b_4)/c_4)^2)$$
$$+ a_5 \exp(-((t-b_5)/c_5)^2) + a_6 \exp(-((t-b_6)/c_6)^2) + a_7 \exp(-((t-b_7)/c_7)^2), \quad\ldots\ldots(5)$$

where

$a_1 = 5109$, $b_1 = 172.5$, $c_1 = 2.314$, $a_2 = 4881$, $b_2 = 182.8$, $c_2 = 1.539$, $a_3 = 14430$, $b_3 = 72.76$, $c_3 = 10.7$, $a_4 = -42800$, $b_4 = 226$, $c_4 = 150.5$, $a_5 = 44280$, $b_5 = 178.9$, $c_5 = 47.44$, $a_6 = 20420$, $b_6 = 102.3$, $c_6 = 24.66$, $a_7 = 15340$, $b_7 = 48.09$, $c_7 = 15.19$.

This is shown in fig.8.

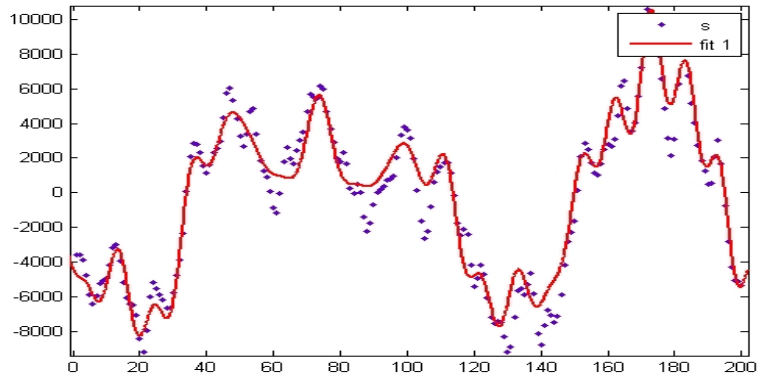

**Fig.8: Best nonlinear curve fit for a smaller segment of the music signal $\{x(k)\}_{k=1}^{10000}$ of raga 'Anandi' of length 200.**

### 6.1.5. Attractor reconstruction of the aforesaid music signal using traditional methods - standard auto-correlation, Average Mutual Information

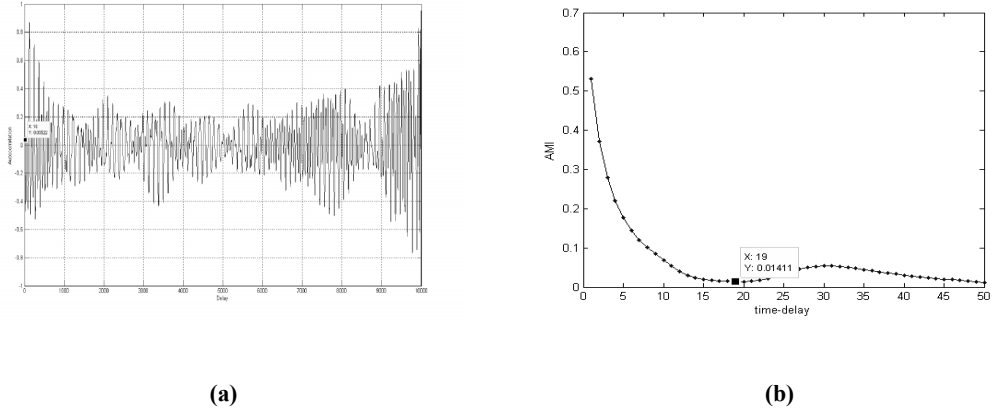

(a)                                                                 (b)

**Fig.9: (a) Plot of auto-correlation against time-delay (*m*) gives *m*=18, (b). Plot of Average mutual information against time-delay (*m*) gives *m*=19.**

From the correlogram diagram- fig.9 (a) and fig.9 (b) under auto-correlation and Mutual Information methods, we get $m = 18$ and $m = 19$ respectively. The corresponding reconstructed attractor is given by fig.10.

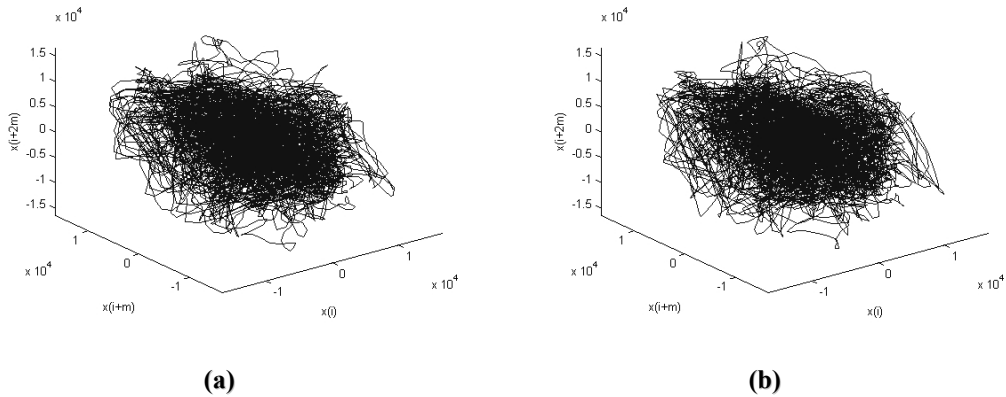

(a)                                                                 (b)

**Fig.10: Reconstructed three dimensional attractor of the music signal of raga 'Anandi' with (a) *m*= 18 under the notion of auto-correlation, (b) *m*= 19 under the notion of Average mutual information.**

### 6.1.6. Attractor reconstruction of the aforesaid music signal using our own methodology

We now apply our own methodology [as discussed in section 2.2.] for the reconstruction of the attractor [17-26] of the aforesaid music signal based on raga 'Anandi'. For this, we use equation (2) to determine the nonlinear auto-correlation $R_X(m)$ between $\{x(k+m)\}_{k=1}^{200-m}$ and $\{f(x(k))\}_{k=1}^{200}$, where $f$ is the best nonlinear fit available for a smaller segment of the present music signal, given by equation (5).

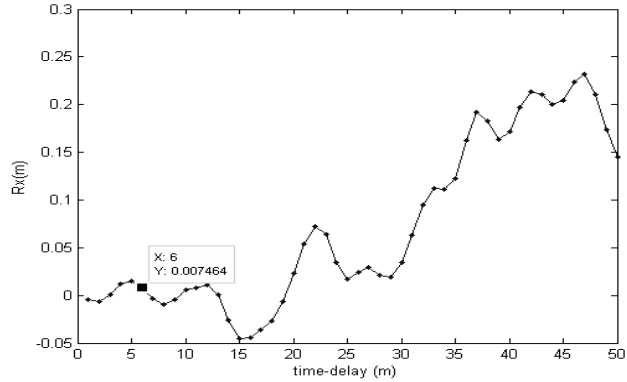

**Fig.11: Two dimensional correlogram diagram for the music signal based on raga 'Anandi'. Here $R_X(m)$ comes nearer to zero for the first time, when $m$=6.**

From the correlogram diagram [32] given by fig.11, we get $m = 6$ and the corresponding reconstructed attractor is given by fig.12.

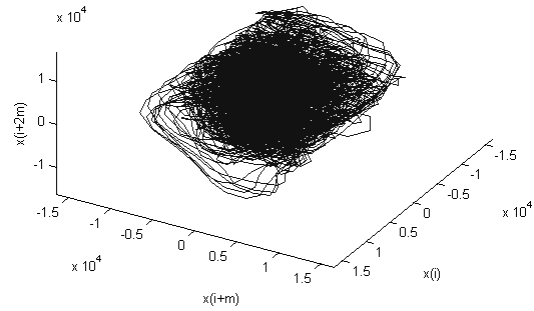

**Fig.12: Reconstructed three dimensional attractor of the music signal based on raga 'Anandi' with $m$= 6 under our newly proposed notion of nonlinear auto-correlation.**

**Remark 6.1**

Three dimensional attractor of fig.10 under standard notion of auto-correlation [31-33] and the notion of Average Mutual Information, shows lack of density in the orbits, and hence they can not be considered as attractors in the proper sense of the term. However the three dimensional attractor reconstructed under our newly proposed methodology, given by fig.12 shows significant improvement over those given by fig.10. In fact, the three dimensional attractor of fig.12 exhibits orbits, which are almost dense except for some outliers. Hence it may be considered as an approximate attractor for the present music signal based on raga 'Anandi'. This claims superiority of attractor reconstruction [17-26] under our newly proposed method of nonlinear auto-correlation over those reconstructed under the traditional methods.

### 6.2. Music signal of a Hindustani Classical Sarod recital based on raga 'Bhairavi'

### 6.2.1. Methodology of collecting data

**Source-** Recorded north Indian Classical Music (Sarod)
**Artist-** Ustad. Amjad Ali Khan
**Raga-** Bhairavi
**Software-** Adobe Audition 1.5
**Samples:** 5000
**Bits per Samples:** 16
**Sample rate:** 44100
**Normalized:** False

### 6.2.2. Test of nonlinearity of the signal

The Surrogate data test [30] with $0.01$ significant level and the statistical parameter AMI $(\tau = 1)$ (Average Mutual Information with time-lag 1) of the present music signal is given in Fig.13.

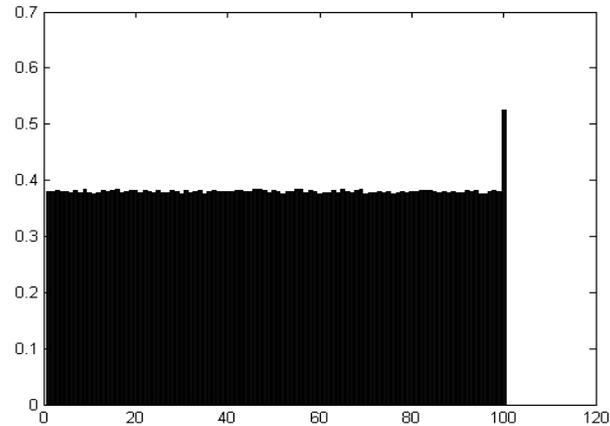

**Fig.13:** Plot of AMI$_{Bhairabi}$ and AMI$_{SUR(Bhairabi)}$ for $\tau = 1$ against signal number, where the first 99 signals are surrogates. Here AMI of the original signal is greater than that of the surrogates. So, the null hypothesis H$_0$: AMI $_{Bhairavi}$ ($\tau = 1$)=AMI $_{SUR\ (Bhairavi)}$ ($\tau = 1$) fails in this case.

Again, we take the null hypothesis $H_0 : AMI_{Bhairavi}(\tau = 1) = AMI_{SUR(Bhairavi)}(\tau = 1)$. If the equality does not hold, we say that null hypothesis fails and alternative $H_A$ holds good. From Fig.13, it is seen that in this case the AMI of surrogate data series is not equal to AMI of the given signal. Hence the null hypothesis $H_0 : AMI_{Bhairavi}(\tau = 1) = AMI_{SUR(Bhairavi)}(\tau = 1)$ is rejected. Thus, nonlinearity of the present signal is established through Surrogate data test [30].

### 6.2.3. Testing for Chaos

It is evident from Surrogate data test [30] of the music signal of raga 'Bhairavi' that the underlying dynamics of the music signal is nonlinear. Also the Lyapunov exponent $(\lambda)$ [27-29] of this music signal is positive. These together imply that the music signal of raga 'Bhairavi' possesses chaotic attractor. So we try for proper attractor reconstruction for this present music signal.

### 6.2.4. Time series plot of the music signal and the best nonlinear curve fit

Let $\{x(k)\}_{k=1}^{5000}$ be 5000 samples of the recorded north Indian Classical Music (Sarod) based on raga 'Bhairavi'. The time series plot of this signal is given by fig.14.

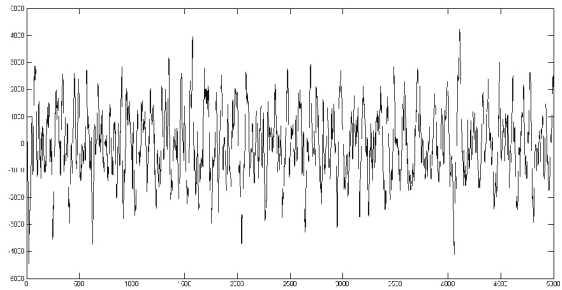

**Fig.14: Time series plot of the music signal based on raga 'Bhairavi'.**

We next consider a smaller segment of the above music signal of length 150 and find a best nonlinear curve fit for it, which is given by

$$f(t) = a_0 + b_1 \sin(wt) + a_2 \cos(2wt) + b_3 \sin(3wt) + a_4 \cos(4wt) + b_5 \sin(5wt) + a_6 \cos(6wt) + b_6 \sin(6wt) + a_7 \cos(7wt) + b_7 \sin(7wt) + a_8 \cos(8wt) + b_8 \sin(8wt), \quad \ldots\ldots(6)$$

where

$a_0 = -5.015 \times 10^{13}$, $b_1 = -4.742 \times 10^{15}$, $a_2 = 8.754 \times 10^{13}$, $b_3 = 4.242 \times 10^{15}$, $a_4 = -6.004 \times 10^{13}$, $b_5 = -4.933 \times 10^{15}$, $a_6 = 4.397 \times 10^{13}$, $b_6 = 4.319 \times 10^{15}$, $a_7 = -2.616 \times 10^{13}$, $b_7 = -1.569 \times 10^{15}$, $a_8 = 4.83 \times 10^{12}$, $b_8 = 2.189 \times 10^{14}$, $w = 0.0816$.

This is shown in fig.15.

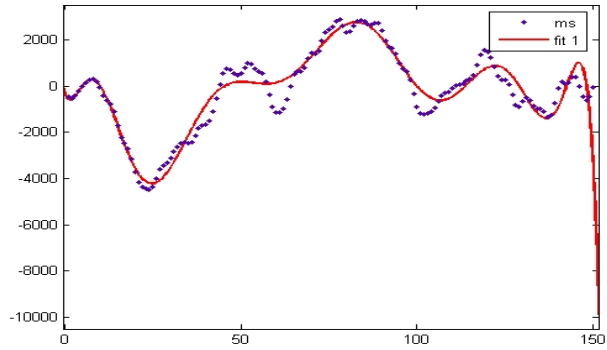

**Fig.15: Best nonlinear curve fit for a shorter segment of the music signal $\{x(k)\}_{k=1}^{5000}$ of raga 'Bhairavi' of length 150.**

### 6.2.5. Attractor reconstruction of the aforesaid music signal using Traditional methods - Minimum auto-correlation, Average Mutual Information

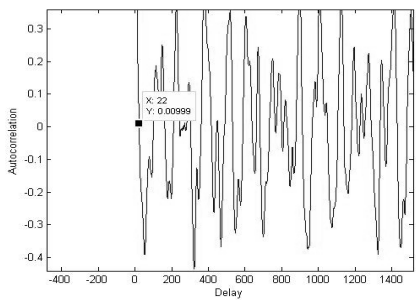
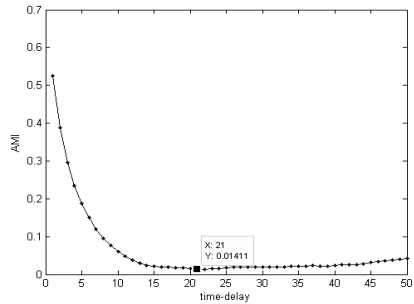

(a)  (b)

**Fig.16: (a) Plot of auto-correlation against time-delay ($m$) gives $m$=22, (b). Plot of Average mutual information against time-delay ($m$) gives $m$=21.**

From the correlogram diagrams - fig.16 (a) and fig.16 (b) under auto-correlation and Mutual Information methods, we get $m = 22$ and $m = 21$ respectively. The corresponding reconstructed attractors are given by fig.17 (a) and fig.17 (b) respectively.

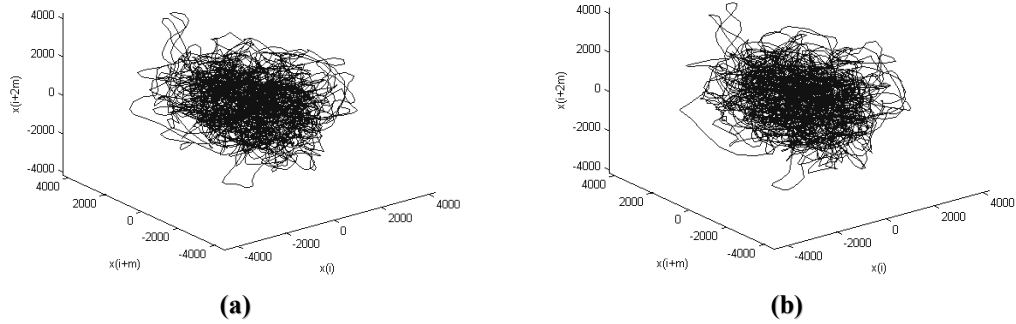

**(a)**                                          **(b)**

**Fig.17: Reconstructed three dimensional attractor of the music signal of raga 'Bhairavi' with (a) *m*= 22 under the notion of auto-correlation, (b) *m*= 21 under the notion of Average mutual information.**

### 6.2.6. Attractor reconstruction of the aforesaid music signal using our own methodology

We now apply our newly proposed methodology [as discussed in section.2.2.] for the reconstruction of the attractor of the aforesaid music signal based on raga 'Bhairavi'. For this, we use equation (6) to determine the nonlinear auto-correlation $R_x(m)$ between $\{x(k+m)\}_{k=1}^{150-m}$ and $\{f(x(k))\}_{k=1}^{150}$, where $f$ is the best nonlinear fit available for a smaller segment of the present music signal, given by equation (6). The values of $R_x(m)$, thus obtained are then plotted against the corresponding values of the time-delay *m* to form the two dimensional correlogram diagram, from where the suitable time-delay *m* is determined.

This is shown in fig.18.

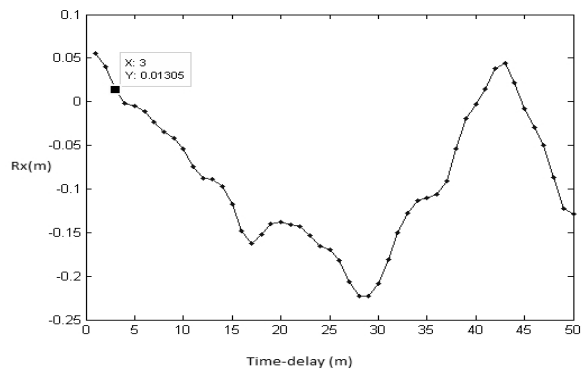

**Fig.18. Two dimensional correlogram diagram for the music signal based on raga 'Bhairavi'. Here $R_x(m)$ comes nearer to zero for the first time when *m*=3.**

From the correlogram diagram given by figure.18, we get $m = 3$ and the corresponding reconstructed attractor is given by figure.19.

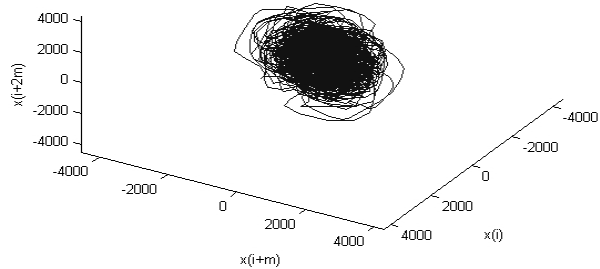

**Fig.19: Reconstructed three dimensional attractor of the music signal based on raga 'Bhairavi' with $m = 3$ under our newly proposed notion of nonlinear auto-correlation.**

**Remark.6.2.**

Three dimensional attractors under standard notion of auto-correlation and the notion of Average Mutual Information given by fig.17 (a) and 17(b) respectively possess many outliers and show lack of density in the orbits and hence they cannot be considered as attractors in the proper sense of the term. But the three dimensional attractor reconstructed under this newly proposed methodology as given by fig.19 is far better than the three dimensional attractors of fig.17 (a) and (b) in the sense that it exhibits orbits, which are almost dense except for very few outliers. Hence it may be considered as the best attractor ever constructed for the present music signal based on raga 'Bhairavi'. This claims superiority of attractor reconstruction under our newly proposed method of nonlinear auto-correlation over those reconstructed under the traditional methods.

**6.3. EMG signal with tremor**

**6.3.1. Methodology of collecting data**

Electromyogram signal (with noise) in analog form of the experimental subjects with tremor were recorded in lead-1 and lead-2 configurations and collected in **data accusation device** available in the **School of BioScience and Engineering, Jadavpur University, Kolkata-32, India** in which it is converted to digital form. This digitized data was then processed in a laptop by using **LAB VIEW** software to remove noise. Finally, the recorded data was analyzed using a **MATLAB** program.

**6.3.2. Testing for Nonlinearity and Chaos**

It is worthy to mention that the underlying dynamics of EMG signal with tremor is nonlinear. This is evident from Surrogate data test [30].

It is also found that the Lyapunov exponent $(\lambda)$ [27-29] of this EMG signal is positive. These together imply that the EMG signal with tremor possesses chaotic attractor. So we try for proper attractor reconstruction for this present EMG signal.

### 6.3.3. Time series plot of the EMG signal and the best nonlinear curve fit

Let $\{x(k)\}_{k=1}^{5000}$ be 5000 samples of the EMG signal with tremor.. The time series plot of this signal is given by fig.20.

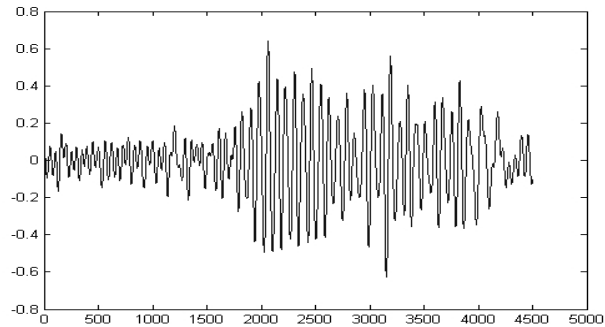

**Fig.20: Time series plot of the EMG signal with tremor.**

We next find a best nonlinear curve fit for a smaller segment of the above EMG signal of length 500, which is given by

$$f(t) = a_1 \sin(b_1 t + c_1) + a_2 \sin(b_2 t + c_2) + a_3 \sin(b_3 t + c_3) + a_4 \sin(b_4 t + c_4) + a_5 \sin(b_5 t + c_5) + a_6 \sin(b_6 t + c_6) + a_7 \sin(b_7 t + c_7) + a_8 \sin(b_8 t + c_8), \dots \dots (7)$$

where
$a_1 = 1.22, \ b_1 = 0.1201, \ c_1 = 2.826, \ a_2 = 0.03432, \ b_2 = 0.06362, c_2 = 2.947,$
$a_3 = 1.191, \ b_3 = 0.1198, \ c_3 = -0.2275, \ a_4 = 0.02456, \ b_4 = 0.02608, c_4 = 2.795,$
$a_5 = 0.03214, \ b_5 = 0.1359, \ c_5 = -1.364, \ a_6 = 0.01691, \ b_6 = 0.0519, c_6 = -1.371.$
$a_7 = 0.01595, \ b_7 = 0.03953, \ c_7 = 0.7243, \ a_8 = 0.01447, \ b_8 = 0.09144, c_8 = -0.9973 \ .$

This is shown in fig.21.

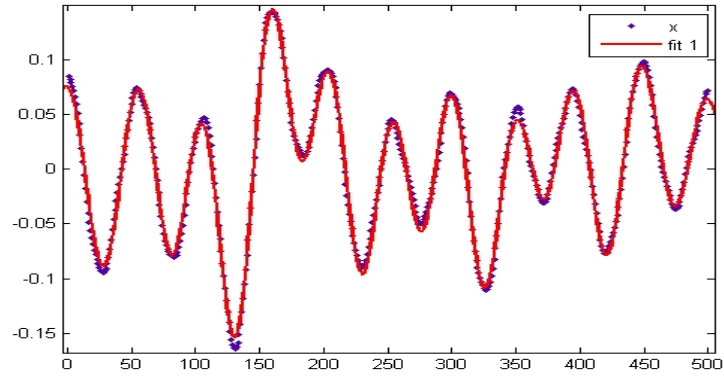

**Fig.21: Best nonlinear curve fit for a shorter segment of the EMG signal $\{x(k)\}_{k=1}^{5000}$ with tremor of length 500.**

### 6.3.4. Attractor reconstruction of the aforesaid EMG signal with tremor using Traditional methods - Minimum auto-correlation, Average Mutual Information

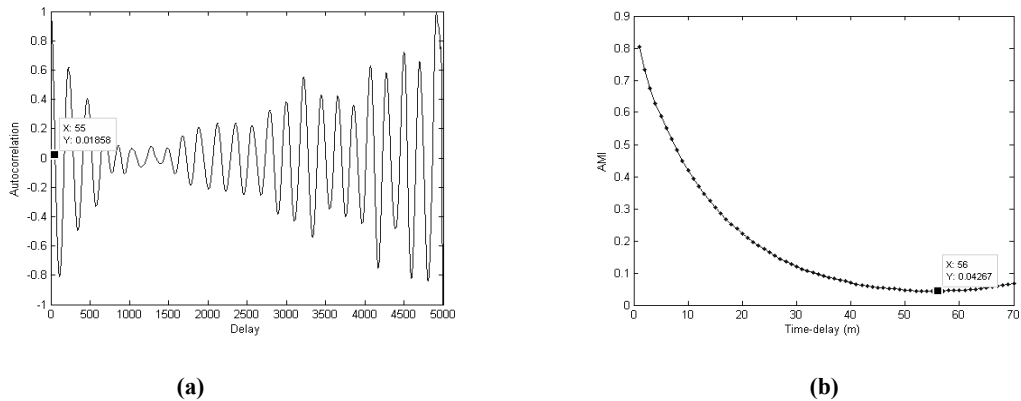

(a)                          (b)

**Fig.22: (a) Plot of auto-correlation against time-delay ($m$) gives $m$=55, (b). Plot of Average mutual information against time-delay ($m$) gives $m$=56.**

From the correlogram diagrams - fig.22 (a) and fig.22 (b) under auto-correlation and Mutual Information methods, we get $m = 55$ and $m = 56$ respectively. The corresponding reconstructed attractors are given by fig.23 (a) and fig.23 (b) respectively.

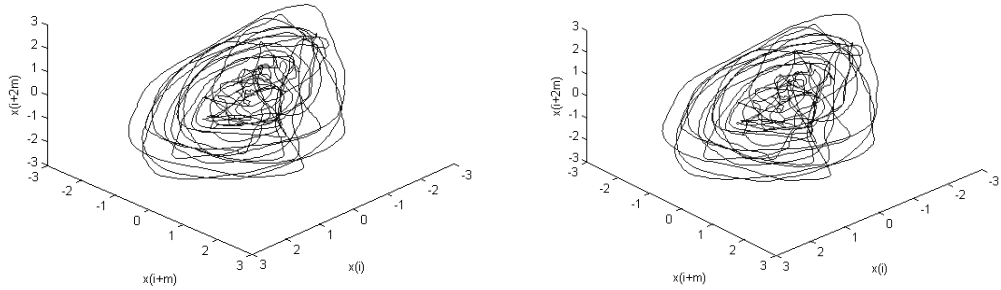

**Fig.23: Three dimensional reconstructed attractor of the above EMG signal with (a) *m*= 55 under the notion of auto-correlation, (b) *m*= 56 under the notion of Average mutual information.**

### 6.3.5. Attractor reconstruction of the aforesaid EMG signal with tremor using our own methodology

We now apply our newly proposed methodology [as discussed in section.2.2.] for the reconstruction of the phase space of the aforesaid EMG signal with tremor. For this, we use equation (7) to determine the nonlinear auto-correlation $R_x(m)$ between $\{x(k+m)\}_{k=1}^{200-m}$ and $\{f(x(k))\}_{k=1}^{200}$, where $f$ is the best nonlinear fit available for a smaller segment of the present EMG signal, given by equation (7). The values of $R_x(m)$, thus obtained are then plotted against the corresponding values of the time-delay *m* to form the two dimensional correlogram diagram, from where the suitable time-delay *m* is determined.

This is shown in fig.24.

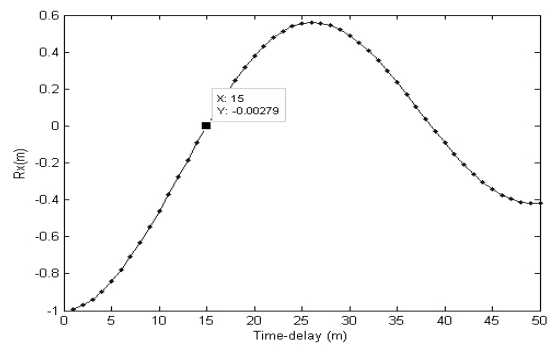

**Fig.24: Two dimensional correlogram diagram for the EMG signal with tremor. Here $R_x(m)$ comes nearer to zero for the first time when *m*=15.**

From the correlogram diagram given by figure.25, we get $m = 15$ and the corresponding three dimensional reconstructed attractor is given by figure.25.

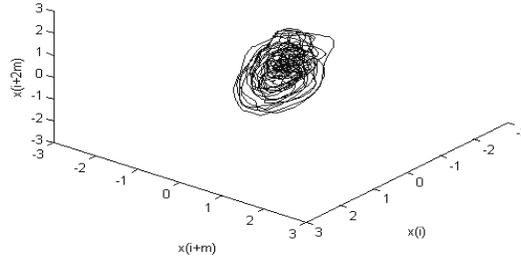

**Fig.25: Three dimensional reconstructed attractor of the above EMG signal with *m*= 15 under our newly proposed notion of nonlinear auto-correlation.**

## Remark.6.3.

It is evident from fig.23 (a) and fig.23 (b) that the three dimensional attractors under the standard notion of auto-correlation and the notion of Average Mutual Information possess many outliers and show lack of density in the orbits. Hence they cannot be considered as attractors in the proper sense of the term. However, the three dimensional attractor reconstructed under this newly proposed methodology as given by fig.25 show much improvement over the previous ones, in the sense that number of outliers gets remarkably decreased in the later case. In fact, the attractor given by fig.25, exhibits orbits that are almost dense except for very few outliers. This again establishes the supremacy of our newly proposed method of nonlinear auto-correlation for attractor reconstruction over the traditional methods.

## 7. Quantifications of the three dimensional Reconstructed attractor

Clustering of points of the reconstructed attractor in three dimensions is a newly proposed quantification technique, which is used to distinguish two different attractors in three dimensions. In this section, we first extend the notion of clustering in two dimensions [40] for any continuous signal with same time-delay.

Let us consider a continuous signal $\{x(k)\}_{k=1}^{N}$ obtained from any system. Also let that the three dimensional attractor is reconstructed by sub-dividing this signal into three groups as $x^+, x^-, x^{--}$ with same delay $\tau$, where

$x^+ = \{x(k)\}_{k=1}^{N-2\tau}$, $x^- = \{x(k)\}_{k=1+\tau}^{N-\tau}$, $x^{--} = \{x(k)\}_{k=1+2\tau}^{N}$,

$1 \leq \tau \leq \dfrac{N}{2}$, if $N$ is even and $1 \leq \tau \leq \dfrac{N-1}{2}$, if $N$ is odd.

This co-ordinate system is transformed by a three dimensional rotation with same angle $\dfrac{\pi}{4}$ ( as the distribution of the points of maximum density on the attractor is along roughly lying with inclination $\dfrac{\pi}{4}$, so we consider the principal axis of the ellipsoid along that line) with respect to X Y and Z axis, which is given by                          .

$$x_m = \frac{1}{2} \cdot x^+ + \left(\frac{1}{2\sqrt{2}} - \frac{1}{2}\right) \cdot x^- + \left(\frac{1}{2\sqrt{2}} + \frac{1}{2}\right) \cdot x^{--} = \frac{2\sqrt{2} \cdot x^+ - \left(\sqrt{2} - 1\right) \cdot x^- + \left(\sqrt{2} + 1\right) \cdot x^{--}}{2\sqrt{2}};$$

$$x_n = \frac{1}{2} \cdot x^+ + \left(\frac{1}{2\sqrt{2}} + \frac{1}{2}\right) \cdot x^- + \left(\frac{1}{2\sqrt{2}} - \frac{1}{2}\right) \cdot x^{--} = \frac{2\sqrt{2} \cdot x^+ + \left(\sqrt{2} + 1\right) \cdot x^- - \left(\sqrt{2} - 1\right) \cdot x^{--}}{2\sqrt{2}};$$

$$x_p = \left(-\frac{1}{\sqrt{2}}\right) \cdot x^+ + \frac{1}{2} \cdot x^- + \frac{1}{2} \cdot x^{--} = \frac{-2 \cdot x^+ + \sqrt{2} \cdot x^- + \sqrt{2} \cdot x^{--}}{2\sqrt{2}}.$$

Thus a new co-ordinate system $(x_m, x_n, x_p)$ is formed.

Let $\overline{x_m} = Mean(x_m), \overline{x_n} = Mean(x_n), \overline{x_p} = Mean(x_p)$ and $SD_1 = \sqrt{Var(x_m)}$, $SD_2 = \sqrt{Var(x_n)}$, $SD_3 = \sqrt{Var(x_p)}$.

Finally, an ellipsoid centered at $(\overline{x_m}, \overline{x_n}, \overline{x_p})$ with three axes of length $SD_1$, $SD_2$ and $SD_3$ is fitted to the existing reconstructed attractor.

As an example, we have applied this quantification technique to quantify the chaotic attractors for the above two music signals and the EMG signal with tremor. The quantification results are summarized in table.1. However, for the purpose of distinction we consider only the above two music signals.

| Reconstructed Attractor of | $SD_1$ | $SD_2$ | $SD_3$ | Volume (Ellipsoid) |
|---|---|---|---|---|
| Signal of a Hindustani Classical music of raga 'Anandi' | 7358.6 | 8153 | 3370.5 | $6.3527 \times 10^{11}$ |
| Signal of a Hindustani Classical music of raga 'Bhairavi' | 2031.7 | 2156 | 686.75 | $9.4545 \times 10^9$ |
| EMG signal with tremor | 1.5689 | 1.5697 | 0.2718 | 2.10255 |

Table.1. Quantification parameters $SD_1$, $SD_2$, $SD_3$ and volume of the three dimensional reconstructed attractors of the above two different music signals and the EMG signal with tremor.

It is observed from table.1 that our newly introduced quantification measure can distinguish the Hindustani classical music of different ragas properly.

## 8. Result and discussion

**i)** The idea of nonlinear auto-correlation introduced in this paper is completely new compared to the earlier ones. In fact this is the first time when attempt has been made to obtain nonlinear auto-correlation according to the type of nonlinearity of the signal. Previously, Average Mutual Information method was applied for the signals irrespective of the nonlinearity of the signals.

**ii)** It gives better result for nonlinear signal analysis compared to the earlier results. Thus it may be claimed that, so far as the reconstruction of three dimensional attractor [17-26] is concerned, possibly this is the best form of the attractor till now.

**iii)** Naturally the interpretation of the signals based on these best forms of the attractors is expected to be most satisfactory.

**iv)** Quantifying parameters for distinguishing the signals are any one of three lengths of the three axes of the ellipsoid, namely $SD_1$, $SD_2$ and $SD_3$ and also its volume V. In fact, all these parameters vary from signal to signal.

**v)** As the music signals are apparently of different types and as they can be distinguished by the musicians properly, so our methodology is tested to be workable in such situations. More precisely, if signals are taken, which are not apparently distinguishable and not even differentiated by the musicians so easily, then also our methodology is of much help in proper distinction of the music signals. Examples are classical Hindustani music compositions based on different ragas, where it is very difficult to establish scientifically that the music compositions are different. But our methodology can be suitably applied in such cases.

**vi)** Our methodology is applicable to any continuous physical signal with different nonlinear trends, not necessarily a music signal alone. It follows from section.6.3, where we have successfully reconstructed the attractor of the continuous EMG signal.